\documentclass{article}

\usepackage{arxiv}

\usepackage[utf8]{inputenc} 
\usepackage[T1]{fontenc}   
\usepackage{hyperref}      
\usepackage{url}           
\usepackage{booktabs}      
\usepackage{amsfonts}      
\usepackage{nicefrac}      
\usepackage{microtype}     
\usepackage{lipsum}

\usepackage{siunitx}
\usepackage{amsmath}
\usepackage{graphicx}
\usepackage[colorinlistoftodos]{todonotes}
\usepackage{tikz}
\usepackage{setspace}
\usepackage{multirow}
\usepackage{acro}
\usepackage{glossaries}
\usepackage{setspace}
\usepackage{caption}
\usepackage{relsize}
\usepackage{array}
\usepackage{standalone}
\usepackage{import}
\usepackage{xcolor}
\usepackage{comment}
\usepackage{soul}
\usepackage{subfig}

\title{TAPsolver: A Python package for the simulation and analysis of TAP reactor experiments}

\author{
  Adam ~Yonge\\
  College of Engineering\\
  Georgia Institute of Technology\\
  Atlanta, GA 30332 \\
  \texttt{ayonge3@gatech.edu} \\
   \And
 M. Ross ~Kunz \\
  Department of Biological and Chemical Processing\\
  Idaho National Laboratory\\
  Idaho Falls, ID 83415 \\
  \texttt{ross.kunz@inl.gov} \\
  \And
 Rakesh ~Batchu\\
  Department of Biological and Chemical Processing\\
  Idaho National Laboratory\\
  Idaho Falls, ID 83415 \\
  \texttt{rakesh.batchu@inl.gov} \\
   \And
 Zongtang ~Fang \\
  Department of Biological and Chemical Processing\\
  Idaho National Laboratory\\
  Idaho Falls, ID 83415 \\
  \texttt{zongtang.fang@inl.gov} \\
   \And
 Tobin ~Issac\\
  College of Computing\\
  Georgia Institute of Technology\\
  Atlanta, GA 30332 \\
  \texttt{tissac@cc.gatech.edu} \\
   \And
 Rebacca ~Fushimi \\
  Department of Biological and Chemical Processing\\
  Idaho National Laboratory\\
  Idaho Falls, ID 83415 \\  \texttt{rebecca.fushimi@inl.gov} \\
  \And
 Andrew J. ~Medford \\
  College of Engineering\\
  Georgia Institute of Technology\\
  Atlanta, GA 30332 \\
  \texttt{ajm@gatech.edu} \\
}

\begin{document}
\maketitle

\begin{abstract}
An open-source, Python-based Temporal Analysis of Products (TAP) reactor simulation and processing program is introduced. TAPsolver utilizes algorithmic differentiation for the calculation of highly accurate derivatives, which are used to perform sensitivity analyses and PDE-constrained optimization. The tool supports constraints to ensure thermodynamic consistency, which can lead to more accurate parameters and assist in mechanism discrimination. The mathematical and structural details of TAPsolver are outlined, as well as validation of the forward and inverse problems against well-studied prototype problems. Benchmarks of the code are presented, and a case study for extracting thermodynamically-consistent kinetic parameters from experimental TAP measurements of CO oxidation on supported platinum particles is presented. TAPsolver will act as a foundation for future development and dissemination of TAP data processing techniques.
\end{abstract}

\keywords{Transient Kinetics \and Algorithmic Differentiation \and Thermodynamic Consistency \and Micro-kinetic Modeling \and Top-Down Analysis}

\section{Introduction}
Catalyst selection and optimization is a necessary step in the design of many chemical production facilities \cite{chorkendorff2017concepts}. Determining an appropriate catalyst for a process, however, is often challenging due to the vast number of variables affecting the catalyst and its environment \cite{medford2018extracting}. Before determining appropriate production-scale operating conditions, the choice of catalyst support \cite{cao2015rational}, composition \cite{galhenage2014understanding}, cluster size \cite{jinnouchi2017predicting,jinnouchi2017extrapolating}, and solvent \cite{saleheen2018liquid} must be considered to optimize product selectivity and yield. Even mild changes to catalyst performance can result in multi-million-dollar profit shifts or substantial changes in pollutant emissions \cite{doi:10.1021/acssuschemeng.7b03652,houdry1959gas}. A robust knowledge of intrinsic kinetic parameters can lead to the rational design of both the catalyst and reactor \cite{matera2019progress}. For these reasons, methods to efficiently probe the intrinsic kinetic properties of catalysts have been, and continue to be, developed \cite{medford2015catmap,hoffmann2014kmos,goldsmith2017automatic,rangarajan2012language}.

Solving for kinetic parameters through reaction model fitting is a standard approach in the field of catalysis \cite{chorkendorff2017concepts,riegel1998kinetic}. The reliability of these fitted parameters, however, are limited by the availability of the experimental data points. For a steady-state experiment, the model is typically sensitive to only a few parameters because the rate-limiting step dominates the extracted kinetic information \cite{caravieilhes2002transient}. This means it is possible to fit multiple kinetic models with comparable quality, leading to uncertainty in the resulting parameters and reaction mechanism \cite{tian2020leveraging, Rangarajan_2017}. Density functional theory (DFT) \cite{kohn1996density} calculations, transition state theory (TST) \cite{laidler1983development} and micro-kinetic modeling (MKM) \cite{dumesicmicrokinetics} are frequently used to address this challenge, leading to fundamental insights about the thermodynamics and kinetics of elementary reactions on catalysts that go beyond what can be deduced from steady-state kinetic experiments. These methods can provide users with information pertaining to the mechanism \cite{mamun2017theoretical}, turnover frequency \cite{nikbin2013dft}, molecular orientation \cite{tereshchuk2014glycerol} and stability \cite{coll2011stability}, but running simulations with DFT requires significant computational resources, even for relatively simple systems \cite{broqvist2002dft,watwe1998density}. Moreover, DFT simulations require an accurate representation of the relevant active sites, which is often unknown for complex catalyst formulations \cite{medford2018extracting}, and the accuracy of the DFT calculations can also be insufficient for quantitatively accurate predictions \cite{medford2014assessing,sutton2016effects,ulissi2011effect}. Surface-science experiments can be used to measure adsorption energies and intrinsic kinetics directly, but they require even more time and resources, and are typically limited to simplified single-crystal surfaces.

\begin{figure}
\centering
\captionsetup{justification=centering}
\usetikzlibrary{shapes.geometric, arrows}
\tikzstyle{letter_circle} = [circle, minimum width=.25cm, text centered, draw=black]
\tikzstyle{startstop} = [rectangle, minimum width=3cm, minimum height=1cm, text centered, draw=black]
\tikzstyle{decision} = [rectangle, minimum width=2cm, minimum height=1cm, text centered, draw=black]
\tikzstyle{process} = [rectangle, minimum width=3cm, minimum height=1cm, text centered, text width=3cm, draw=black]
\tikzstyle{arrow} = [thick,->,>=stealth]
\begin{tikzpicture}[node distance=2cm,scale=0.77,every node/.style={scale=0.77}]

<TikZ code>
\draw (-3,1) -- (14,1) -- (14,-9) -- (-3,-9) -- cycle;
\draw (-2,1) -- (-2,-9);
\draw [thick,dashed] (-3,-1) -- (14,-1);
\draw [thick,dashed] (-3,-3) -- (14,-3);
\draw [thick,dashed] (-3,-5) -- (14,-5);
\draw [thick,dashed] (-3,-7) -- (14,-7);
\node (start) [startstop] {\textbf{Gas Species}};
\node (act1) [process, right of=start, xshift = 2cm] {\textbf{Mechanism \& Parameters}};
\node (step2) [process, right of=act1, xshift = 2cm] {\textbf{Experimental Data}};
\node (step3) [process, below of=step2] {\textbf{Parameter Fitting}};
\node (step8) [process, right of=step3, xshift=2cm] {\textbf{Sensitivity Analysis}};
\node (step9) [process, below of=step8] {\textbf{Parameter Sensitivities}};
\node (step7) [process, below of=step3] {\textbf{Kinetic Parameters}};
\node (step4) [process, below of= start] {\textbf{Mechanism Generator}};
\node (step5) [process, right of=step4, xshift=2cm] {\textbf{Pulse Simulator}};
\node (step6) [process, below of=step5] {\textbf{Synthetic Data}};
\node (step10) [process, below of=step6] {\textbf{Y/G Procedure}};
\node (step14) [process, left of=step6, xshift=-2cm] {\textbf{Experimental Data}};
\node (step11) [process, right of=step10, xshift=2cm] {\textbf{Rate-Reactivity Model}};
\node (step12) [process, below of=step10] {\textbf{Rate/Concentration in Catalytic Zone}};
\node (step13) [process, below of=step11] {\textbf{Reactivities}};

\draw (-2.5, 0) node {\rotatebox{90}{\textbf{\underline{Inputs}}}};
\draw (-2.7, -2) node {\rotatebox{90}{\textbf{\underline{Upstream}}}};
\draw (-2.3, -2) node {\rotatebox{90}{\textbf{\underline{Processes}}}};
\draw (-2.7, -4) node {\rotatebox{90}{\textbf{\underline{Initial}}}};
\draw (-2.3, -4) node {\rotatebox{90}{\textbf{\underline{Output}}}};
\draw (-2.7, -6) node {\rotatebox{90}{\textbf{\underline{Downstream}}}};
\draw (-2.3, -6) node {\rotatebox{90}{\textbf{\underline{Processes}}}};
\draw (-2.7, -8) node {\rotatebox{90}{\textbf{\underline{Downstream}}}};
\draw (-2.3, -8) node {\rotatebox{90}{\textbf{\underline{Output}}}};
\draw [arrow] (step5) -- (step3);
\draw [arrow] (step2) -- (step3);
\draw [arrow] (step8) -- (step9);
\draw [arrow] (step3) -- (step8);
\draw [arrow] (step3) -- (step7);
\draw [arrow] (step5) -- (step6);

\draw [arrow] (step6) -- (step10);
\draw [arrow] (step10) -- (step11);
\draw [arrow] (step10) -- (step12);
\draw [arrow] (step11) -- (step13);

\draw [arrow] (start) -- (step4);
\draw [arrow] (step4) -- (step5);
\draw [arrow] (act1) -- (step5);

\draw [arrow] (step14) |- (step10);

\end{tikzpicture}
\caption[Workflow for Processing TAP Reactor Pulses]{Workflow highlighting traditional and recently developed methods of processing TAP reactor experiments, with upstream processes representing analyses that involve preliminary mechanistic assumptions and downstream processes excluding them. The methods utilized by TAP experimentalists are likely to expand and access to new approaches must be streamlined.} \label{fig-flwdgrm-2}
\end{figure}
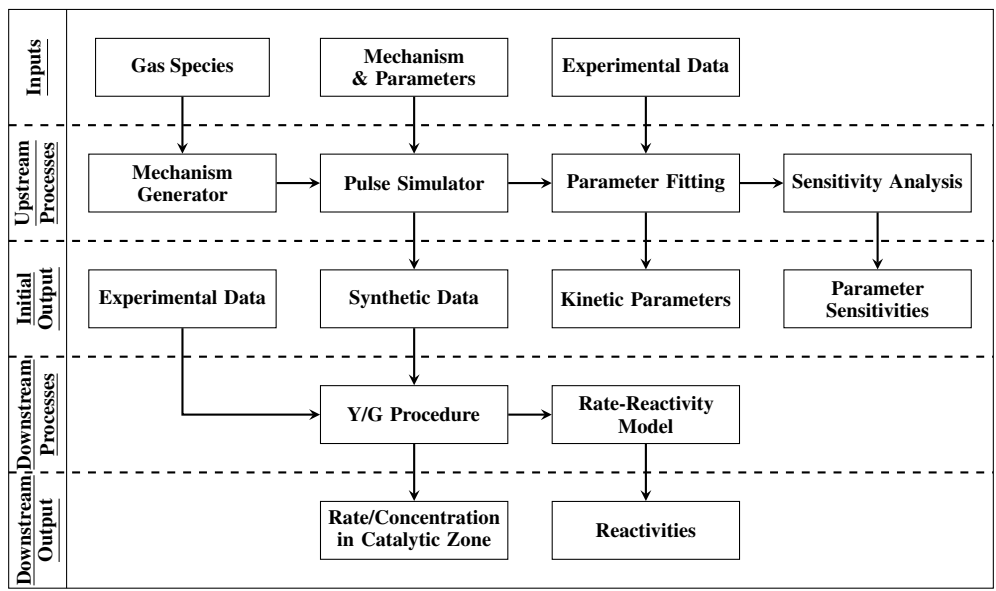

Transient kinetic experiments are an additional technique that can complement steady-state experiments, computational chemistry, and traditional surface science experiments \cite{moncada2018developing}. Transient experiments can provide users with a nuanced understanding of the intrinsic kinetics of a system \cite{biloen1983transient}, but effectively decoupling the transport and kinetics is often complex \cite{yablonsky2016rate}. The temporal analysis of products (TAP) reactor system was developed to provide well-defined transport properties and enable kinetic studies of complex supported catalysts. TAP experiments consist of a series of nano-mole pulses operating at under isothermal conditions within the Knudsen diffusion regime, and data are collected at a millisecond time resolution \cite{morgan2017forty,gleaves1988temporal}. This provides rich and detailed insight into the transient kinetic response of supported catalysts. However, a noteable challenge of working with TAP data is the complex data analysis required to convert the raw data to meaningful kinetic information. Steps have been taken through the years to account for these issues, with the development of the governing partial differential equations (PDEs) \cite{kondratenko2008mechanistic,kondratenko2010mechanism,rothaemel1996modeling,kumar2011microkinetic}, methods to solve them, and examples of solutions for different transient experiments \cite{van1997mathematical}. This has provided a foundation for prior TAP simulators \cite{menon2011reaction,roelant2011mathematical,balcaen2011kinetic,batchu2019ethanol,delgado2002modeling,reece2017kinetic}. The number of methods have expanded over the years, as presented in Figure \ref{fig-flwdgrm-2}), with the development of approaches for the construction of mechanisms, simulation of pulses, fitting of parameters and data-driven models for analyzing results  \cite{kunz2018pulse,yablonsky2016rate}.

Although several TAP analysis packages exist in the field, some obstacles still remain. Most codes are not easily or publicly accessible, instead requiring users to contact developers directly to gain access to code, and many make use of proprietary frameworks or programming languages that require costly licences. In addition, validation and documentation can be sparse or convoluted, making it difficult to confirm the accuracy of results without frequent discussion with the original developers. 

In the case of TAP fitting programs, the optimization algorithms and loss functions are often hard-coded, making it difficult to implement emerging optimization techniques from machine learning. The TAP fitting codes also require strong initial guesses and do not provide estimates of uncertainty, making it challenging to draw strong conclusions about the fitted parameters or mechanisms. 
To address these issues, the Python program TAPsolver, a TAP reactor simulation and analysis tool, is developed and presented. This program will be made freely available for academic research. TAPsolver is built around the open-source FEniCS \cite{alnaes2015fenics,logg2012automated,logg2012ffc} and Dolfin-Adjoint \cite{LoggWellsEtAl2012a,logg2010dolfin,mitusch2019dolfin} packages, which solve PDEs via finite element methods and utilize model adjoints to enable algorithmic differentiation. TAPSolver is designed to be compatible with external optimization routines, enabling experimentation with various algorithms, and constraints can easily be added to the loss function or optimization algorithm to enforce thermodynamic consistency or other known information.
TAPSolver flexibly accepts different reaction mechanisms and rate/reactor parameters, and interfaces with other techniques for TAP data analysis, making it a natural choice integrating future developments in TAP data analysis.
In this work, the commands used to define TAPsolver data processing are introduced and TAPsolver is applied to both synthetic and experimental data. 

\section{Mathematical Background}

The fundamental PDEs and boundary conditions used for TAP simulations and constraints applied to optimize rate constants and enforce thermodynamic consistency
are briefly outlined in this section. 
More rigorous details can be found in publications by Van der Linde et al. \cite{van1997mathematical}, Farrell et al. \cite{farrell2013automated} and Mhadeshwar et al. \cite{mhadeshwar2003thermodynamic}, or elsewhere \cite{griewank1989automatic,martins2001connection}.

\subsection{Defining the Transport and Reaction Equations}\label{transportEqs}

Mathematical descriptions of catalytic experiments are often complex, involving multidimensional heat, mass and momentum transfer at different temporal and spatial scales. Current limitations of computational resources often force researchers to balance simplifying assumptions with accuracy. TAP reactors, however, are setup to reduce many of these complexities. TAP primarily operates in the Knudsen diffusion regime and consists of two equally sized inert zones surrounding a thin catalytic region. Though reactions in the catalyst zone make the system inherently complex and can lead to challenging simulations, there are simplifications that can limit this burden. It has been shown that a one-dimensional model is acceptable when the length of the TAP micro-reactor is at least three and a half times the radius of the reactor \cite{constales2001multi}. This drastically reduces the complexity of the equations and the associated computational expense of solving them. The PDE describing the diffusion of a gas species through a TAP reactor, with a general reaction term, is

\begin{equation}\label{PDE-form-gas}
	 \varepsilon{}_{j}\frac{\partial C_{i}(x,t)}{\partial t} - \frac{\partial}{\partial x}\cdot{}(D_{i,j}\;{}\nabla{}C_{i}(x,t)) = R(C_{i}(x,t),\theta_{i}(x,t),k_{j},...)
\end{equation}

where $C_{i}(x,t)$ represents the concentration of gas species $i$, $\varepsilon{}_{j}$ is the void fraction of the material in zone $j$, $t$ is time, $x$ is the length of the reactor, $\theta_{i}$ are surface intermediates $i$ and $D_{i,j}$ is the Knudsen diffusion coefficient of gas $i$ in zone $j$. The change in surface intermediates with time is defined as 

\begin{equation}\label{PDE-form-surf}
	 \frac{\partial \theta_{i}(x,t)}{\partial t} = R_i(C_{j}(x,t),\theta_{j}(x,t),k_{l},...)
\end{equation}

The reaction term $R$ is the microkinetic model, defined as \cite{gleaves2010temporal}

\begin{equation}\label{reaction-expression}
    \ R_{i} = \sum_m s_{im} \left(k_{m}^{+} \prod_{j \in F_m} \theta_{j}(x,t) \prod_{j \in F_m} C_{j}(x,t) - k_{m}^{-} \prod_{l \in B_m} \theta_{l}(x,t) \prod_{l \in B_m} C_{l}(x,t)\right)
\end{equation}

where $s_{im}$ is the stoichiometry of species $i$ in elementary step $m$, $k_{m}^{+/-}$ are the forward/reverse rate constant for elementary step $m$, and $F_m$ and $B_m$ are sets of indices corresponding to the intermediate and gas-phase species in the forward and backward reactions of elementary step $i$. Solving this PDE requires specifying  boundary conditions. In the case of the TAP experiment, the initial condition of gas species are

\begin{equation}
	 C_{g}(x, t_{p}) = \delta{}(x, t_{p})
\end{equation}

where $\delta{}(x, t_{p})$ is a delta function 
introduced at the pulse time $t_{p}$. The initial condition of surface intermediates inside the catalyst zone is

\begin{equation}
	 \theta{}(P_{1} < x < P_{2}, t_{0}) = U_{i}
\end{equation}

where $U_{i}$ is the initial surface concentration of adsorbate or active site $i$ and $P_{1}$ and $P_{2}$ represent the normalized domain of the catalyst zone in the reactor. These values can be changed based on the fraction the reactor occupied by the catalyst. The initial condition of surface intermediates outside the catalyst zone is

\begin{equation}
	 \theta{}(x < P_{1} \ and \ x > P_{2}, t_{0}) = 0
\end{equation}

 where no surface concentration is considered due to assumptions of inert interactions with non-catalytic materials. At the entrance of the reactor, a Neumann boundary condition can be imposed due to the use of a pulse valve and is written as

\begin{equation}
	 \frac{\partial C_{g}(0,t_{p})}{\partial x} = 0
\end{equation}

while a vacuum is being applied at the exit of the reactor, leading to the Dirichlet boundary condition

\begin{equation}
	 C_{g}(L, t) = 0
\end{equation}

with $L$ being the length of the reactor. Multiple species can be introduced into the reactor system simultaneously, indicated by the index $g$. When a species is formed in the catalyst region but not pulsed into the reactor, then the initial intensity of the Dirac delta function associated with that species is set to zero. All gas species will have a concentration of zero at the exit of the reactor due to the applied vacuum conditions, and the outlet flux is defined as 

\begin{equation}
    F_{i} = D\frac{\partial C_{i}(x,t)}{\partial x}_{x=L}
\end{equation}

The corresponding variational form for the transport term of Equation \ref{PDE-form-gas}, needed for solving the PDEs using finite-element methods, is written as \cite{logg2012automated}

\begin{equation}\label{variational_non_reaction}
	\int \bigg( \frac{\varepsilon{}_{j}(C_{i}^{n+1}-C_{i}^{n})v_{i}}{\triangle{}t} + D_{i,j} \nabla{}c_{i}^{n+1}\cdot{}\;{}\nabla{}v_{i} \bigg) \;dx = 0
\end{equation}

where $C_{i}^{n+1}$ and $C_{i}^{n}$ are the species concentrations at current and previous time steps, respectively, $\textit{$\Delta$t}$ is the size of the time step and $\textit{v$_{i}$}$ is the test function for each of the gas species. The terms test and trial functions are frequently used in variational problem notation and represent the components for performing integration by parts over the spatial domain. Reactive terms can be added to the right-hand side of Equation \ref{variational_non_reaction} and consist of combinations of the concentration terms. As a specific example, consider the simple gas-phase reaction

\begin{equation}
    A \rightarrow{} B
\end{equation}

with a rate constant $k$.  The kinetic model is defined as

\begin{equation}
    kC_{A}v_{A}dx - kC_{A}v_{B}dx = 0
\end{equation}

where each term represents the consumption and generation of species $A$ and $B$, respectively. When combined, the variational form of the reaction-diffusion equation is defined as

\begin{equation}
\begin{split}
    \int \bigg( \frac{(C_{A}^{n+1}-C_{A}^{n})v_{A}}{\triangle{}t} + D_{A} \nabla{}c_{A}^{n+1}\cdot{}\;{}\nabla{}v_{A} \bigg) \;dx \\
    & + \int \bigg( \frac{(C_{B}^{n+1}-C_{B}^{n})v_{B}}{\triangle{}t} + D_{B} \nabla{}c_{B}^{n+1}\cdot{}\;{}\nabla{}v_{B} \bigg) \;dx = kC_{A}v_{A}dx - kC_{A}v_{B}dx
\end{split}
\end{equation}

\subsection{Curve Fitting and Thermodynamic Consistency}\label{objectiveSection}

One key advantage of TAP data is that the TAP curve will implicitly contain information about multiple elementary steps, and TAP curve fitting has been commonly used to extract intrinsic kinetic parameters from TAP data \cite{menon2011reaction,roelant2011mathematical,balcaen2011kinetic,batchu2019ethanol,delgado2002modeling,reece2017kinetic}. Fitting TAP curves to extract rate constants is a form of PDE-constrained optimization, and requires an objective function to optimize. 
Agreement between model and experiment can be optimized by using an objective function corresponding to the summation of the difference between each of the simulated and experimental outlet flux data points, defined as

\begin{equation}\label{objectiv_function}
	J_{data} = \mathlarger{\mathlarger{\sum_{i}}} ( \frac{1}{2}D \int_{L-\triangle{}x}^{L} | u_{i}(x,t) - u_{i,obs}(x,t)|^{2} dx ) 
\end{equation}

where $i$ represents each of the points considered, $L$ is the length of the reactor, $u_{i}$ is the outlet concentration of the forward solution, $u_{i,obs}$ is the experimental outlet concentration, and $\triangle{}x$ is the mesh step size. Since the experimental data corresponds to the outlet flux, the inner product is evaluated over the spatial step at the reactor outlet, i.e. $L - \triangle{}x$ to $L$. Other objective functions could be defined that involve additional experimental data or observable values (e.g. surface species, spectroscopic information, integral quantities) \cite{savara2016simulation}. Since outlet fluxes are the most common observed and fitted values for TAP experiments, this provides a natural starting point.

In addition to agreement with the data, it is also important to consider the thermochemistry of the reaction. Any set of kinetic parameters that does not obey the proper thermodynamic constraints will lead deviations in the predicted equilibrium concentrations and errors with the energy balance will arise \cite{salciccioli2011review,motagamwala2018microkinetic}. In addition, Equation \ref{objectiv_function} alone may not always be adequate for identifying the appropriate reaction mechanisms or kinetics, so the additional constraint may aid in mechanism discrimination. For example, micro-kinetic analysis through DFT and fitting results of steady-state kinetic experiments often utilizes thermodynamic consistency constraints. \cite{aghalayam2000construction,mhadeshwar2003thermodynamic} These approaches have been used to help test the validity of many studies, including the partial oxidation of methane \cite{mhadeshwar2005hierarchical}, ethylene hydrogenation \cite{salciccioli2011microkinetic} and the water-gas shift reaction \cite{grabow2008mechanism}.

\begin{figure}[bt]
\centering
\captionsetup{justification=centering}
\includegraphics[width=7cm]{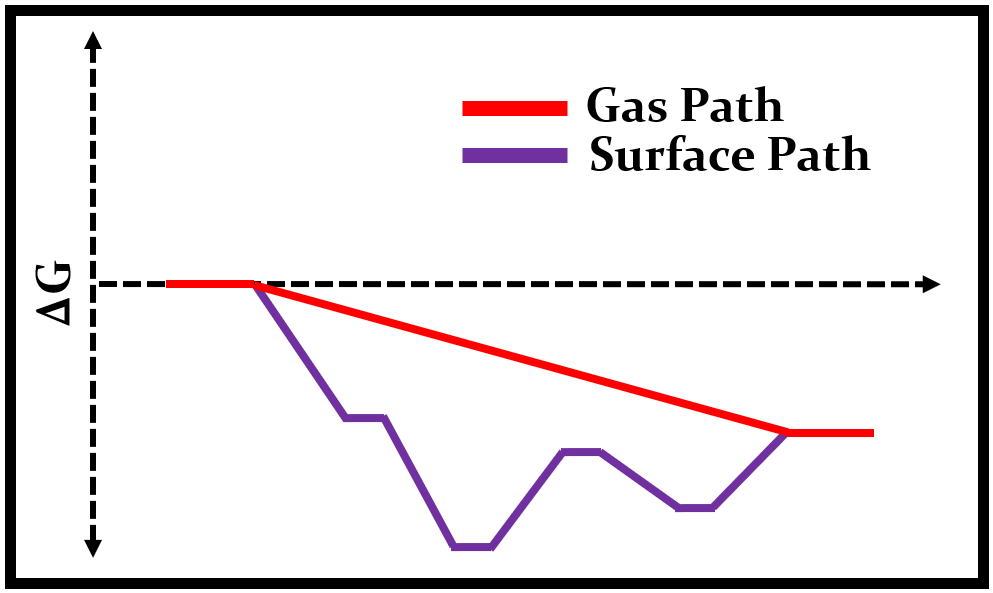}
\caption{\label{thermoCons} Free energy diagram illustrating thermodynamic consistency that can be enforced during PDE-constrained optimization with TAPSolver. The summation of elementary free energies of reaction (purple) must match the gas phase free energy of reaction (red).}
\end{figure}

Variations of the thermodynamic consistency equations have been outlined by Dumesic et al. \cite{dumesic1993microkinetics} Although the implementation of each form of thermodynamic consistency would be useful, many forms require the definition or estimation of intermediate values (i.e. the thermodynamic data per elementary processes) that may not be known. However, the thermochemistry of the overall reaction, visualized in Figure \ref{thermoCons}, is often well-known, so it is useful to be able to ensure that the optimum reaction parameters are consistent with this information, which will be independent of the catalyst material used.

The free energy of reaction is defined as 

\begin{equation}
	\Delta{G_{gas}} = \Delta{H} - T*\Delta{S}
\end{equation}

where the enthalpic and entropic terms for individual gas species are summed to define $H$ and $S$, respectively. To determine the contributions of individual elementary reactions (denoted $i$), the equilibrium constant is defined as 

\begin{equation}
	 K_{i} = \frac{k_{f,i}}{k_{b,i}}
\end{equation}

and the free energy of the process is calculated through

\begin{equation}
	 \Delta{G_{i}} = -R*T\ln{(K_{i})}
\end{equation}

with $R$ representing the ideal gas constant and $T$ the reactor temperature. The difference between the free energy of reaction and each of the elementary free energies is the additional term of the objective function defined in Equation \ref{objectiv_function} to account for thermodynamic consistency, written as 

\begin{equation}
	 J_{thermo} = (\Delta{G_{gas}} - \sum_{i \in M}\Delta{G_{i}})^{2}
\end{equation}

where $M$ is the set of elementary steps for a given serial mechanism corresponding to the gas-phase reaction.
A combined objective function, given by 

\begin{equation}
	 J = J_{data} + \alpha J_{thermo}
\end{equation}

can then be utilized to simultaneously fit experimental TAP data and satisfy thermodynamic constraints, with the parameter $\alpha$ controlling the relative importance of thermodynamics compared to data.
Thermodynamics have been considered alongside TAP reactor studies previously \cite{yablonsky2014decoding,beck2014oxidative}, but as an additional component in interpreting results rather than constraining optimized parameters. To our knowledge, no other TAP optimization code supports thermodynamic consistency constraints. 

\section{TAPsolver Implementation}\label{tapImplem}

The FEniCS package is an established code for efficient finite element simulation, and the associated Dolfin-Adjoint package utilizes model adjoints to provide algorithmic differentiation of the models. One challenge in TAP simulations is that significant modifications to the underlying PDEs are required to simulate different microkinetic mechanisms. 
Manually changing these equations for different experimental setups or chemical systems is time consuming and error-prone. 
However, automating the construction of these equations provides an efficient route to making FEniCS practical for TAP reactor analysis. Similarly, having to manipulate code directly to set initial conditions (e.g. pulse intensities, surface concentrations, etc.) or analysis methods (e.g. curve fitting, sensitivity analysis) can also become a bottleneck.  To overcome these issues, TAPsolver has a spreadsheet-based input file to simplify the construction of these equations, and the code is modular, facilitating facile access to various analysis methods. This section outlines each of the components of the input file and the available analysis techniques. 

\begin{table}

\begin{center}

 \begin{tabular}{|c|c|c|c|} 
 %\hline
 \hline
 \textbf{Reactor Setup} & Zone 1 & Zone 2 & Zone 3 \\ %&  \textbf{units}
 \hline
 Zone Length & 3.0 & 0.1 & 2.9 \\
 \hline
 Zone Void & 0.4 & 0.4 & 0.4 \\
 \hline 
 Reactor Radius & 1 &  &  \\
 \hline 
 Reactor Temperature & 400 &  &  \\
 \hline  
 Mesh Size & 200 &  &  \\
 \hline   
 Catalyst Mesh Density & 4 &  &  \\
 \hline   
 Output Folder & results &  &  \\
 \hline
 Experimental Data Folder & ../data &  &  \\
 \hline    
 Reference Diffusion Inert & 13.5 &  &  \\
 \hline    
  Reference Diffusion Catalyst & 13.5 &  &  \\
 \hline    
  Reference Temperature & 13.5 &  &  \\
 \hline    
  Reference Mass & 40 &  &  \\
 \hline
  &  &  &  \\
 \hline\hline\hline
 \textbf{Feed and Surface Composition} &  &  &  \\
 \hline
  & $CO$ & $O_{2}$ & $CO_{2}$ \\
 \hline
 Intensity & 5 & 5 & 5 \\
 \hline 
 Time & 0 & 0 & 0 \\
 \hline 
 Mass & 28 & 32 & 44 \\
 \hline  
  & & & \\
 \hline   
  & CO* & O* & * \\
 \hline
 Initial Composition & 0 & 12 & 12 \\
 \hline
  & & & \\
 \hline\hline\hline
 \textbf{Elementary Reactions} &  &  &  \\
 \hline
 $CO + * <-> CO*$ & $1*10^{-10}$ & $1*10^{-10}$ &  \\
 \hline
 $O2 + 2* <-> 2O*$ & $1*10^{-10}$ & $1*10^{-10}$ &  \\
 \hline
 $CO* + O* <-> CO2 + 2*$ & $1*10^{-10}$ & $1*10^{-10}$ &  \\%& ------------------------ \\ [1ex]
 \hline\hline\hline 
 \textbf{Thermodynamic Consistency} & & & \\
 \hline
 r1 + (0.5)*r2 + r3 &  & & \\
 \hline
 \end{tabular}
\end{center}
\caption[Example Input CSV/XLS File Used to Simulate TAP Pulses]{\label{usr_inp} An example of the input file used for TAPsolver, consisting of reactor dimensions, initial conditions, elementary reactions and optional thermodynamic consistency constraints.}
\end{table}

\subsection{Defining Reactor Setup, Feed and Surface Composition}

The reactor setup, feed and surface composition, and reaction mechanism specification are the primary components of the TAPsolver input file, presented in Table \ref{usr_inp}, with thermodynamic consistency being an optional feature. Many of the variables defined in the input file are repeatedly used, enabling files to be copied and minimally modified. A brief discussion of the key inputs is provided here, and more details are available in the documentation.

Several parameters are used to define the layout of the simulated TAP reactor, including the reactor length, radius, void fraction, catalyst location and fraction, and temperature. For operation in the Knudsen transport regime, a reference diffusion must be defined, including the reference temperature and mass in both the inert and catalyst domains. This information is used to determine the diffusion coefficient for all gaseous species based on the mass of each through scaling \cite{mason1967graham}.
In addition to variables used to define the reactor setup are the variables that to control the simulation accuracy. The primary accuracy variable is the the 'Mesh Size' which defines the spatial precision.
TAPsolver also supports mesh refinement within the catalyst domain. Simulating transport through inert zones requires a much coarser mesh than the catalyst domain where reactions happen. 
The 'Catalyst Mesh Density' input parameter refines the precision of the mesh in the catalyst zone while keeping the mesh in both inert zones constant. For example, a mesh size to simulate transport in a catalyst free reactor would be approximately two hundred cells. If the user wanted to run a simulation with fifty cells within the catalyst zone that occupied 2\% of the reactor, the simulation would require a uniform mesh size of 2500 cells. With the mesh density parameter, it is possible to have a catalyst mesh size of sixty-four (with the initial four cells in the catalyst zone being doubled four times), while the total reactor mesh size is 264. This can substantially improve accuracy with a minimal impact on simulation time.

The reactant pulse intensity, pulse time, mass, and initial surface composition are defined in the input file below their associated gas names. 
The user can also specify inert gas pulses that are not involved in the reaction process to the right of reactive gas species. Pulses can also be introduced at later times, enabling the simulation of multi-pulse or ``pump-probe'' experiments. The 'Output Folder' and 'Experimental Data Folder' variables specify the file structure of analysis results and directly link the experimental data to the simulations, respectively. A series of nested folders are generated following each simulation that contain all data for the processes specified by the user. 
'thin data', 'sensitivity analysis' and 'uncertainty quantification'. 
The input file used to run the simulation is also stored in the output folder to improve data veracity.

\subsection{Defining Elementary Reactions in TAPsolver}

The set of elementary reactions and list of gaseous species, adsorbed species, and active sites defined by the user are converted into a matrix defining the microkinetic model \cite{gusmao2015general} and subsequently used to produce the variational form that is solved by FEniCS. 
The elementary reactions defined in the input file are flexible, with reversible and irreversible reactions being defined through the arrow (either a reversible or right pointing arrow) and the values of the kinetic parameters being placed in the cells to the right of the elementary reactions, with the first cell representing the forward rate constant and the second cell representing the reverse rate constant. Three options are available for the definition of kinetic parameters in the input file, including direct rate constants, pre-exponential/activation enthalpy (noted with a ''\$'' symbol), or activation free energy (noted with a ''@'' symbol). Users can also select which of these kinetic parameters to include or exclude in the optimization routine through the inclusion of an exclamation point following the value of the parameter, making it easier to target specific values of interest. Examples of the previous parameter options are shown in Table \ref{exampleReactions}, with a reversible, irreversible, activation enthalpy based, free energy based and fixed parameter values being presented through each elementary reaction, respectively.   

\begin{center}
\small\addtolength{\tabcolsep}{-5pt}
\begin{longtable}{|c|c|c|}
\caption{Examples of how to define elementary reactions and their kinetic parameters for use by the TAPsolver input file.} \label{exampleReactions} \\
\hline Elementary Reaction &  Forward & Reverse \\
\endfirsthead
\hline  $A + * \leftrightarrow A^{*}$ & 1
 & 2 \\
\hline  $B + * \rightarrow B^{*}$  & 1
 & -- \\
\hline  $A* + B* \rightarrow C^{*}$  & 1\${}2 & -- \\
\hline $A* + B* \rightarrow A^{*}$ & 1@2 & -- \\
\hline $C^{*} \rightarrow C + *$ & 1! & -- \\
\hline
\end{longtable}
\end{center}
\subsection{Running Analyses}

\begin{figure}[bt]
\centering
\captionsetup{justification=centering}
\frame{\includegraphics[width=11cm]{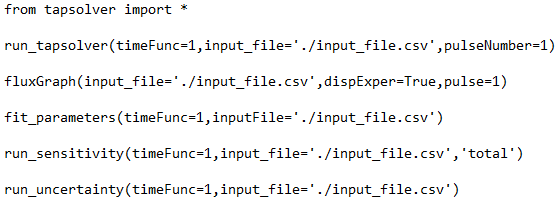}}
\caption{An example code snippet showing how TAPsolver is implemented in Python, with available functions for running a simulation, displaying graphs, fitting parameters, and evaluating parameter sensitivities (the gradient).}\label{codeSnippet}
\end{figure}

TAPsolver is structured to enable easy access to various features and analysis routines via Python functions.
A snippet of TAPsolver code is presented in Figure \ref{codeSnippet}, with examples implementations of some implemented features. Forward simulations of TAP pulses can be performed with the \texttt{run\_tapsolver} command along with the name of the input file, the simulation time (\texttt{timeFunc}) and the number of pulses (\texttt{pulseNumber}). Once the simulation finishes the resulting outlet fluxes can be plotted using the \texttt{fluxGraph} function. 
Desired subsections of the simulated and experimental data can be specified with the parameter 'pulse', making it easier to observe trends and visualize results along with experimental data. Methods related to the inverse problem can also be programatically utilized. Parameters can be fitted with the \texttt{fit\_parameters} with similar inputs to the forward problem.  
The gradient of the objective function with respect to fitted parameters can also be calculated through the \texttt{run\_sensitivity} command. One feature of the sensitivity analysis is the option to compute the sensitivity over time (the influence of each kinetic parameter with respect to the objective function at each time step), as opposed to being defined in terms of the total objective function (used in optimization). This can be achieved using the sensType argument in \texttt{run\_sensitivity}, and enables the user to identify which parts of the TAP curve are controlled by specific elementary steps.

\section{Results and Discussion}

Both the forward and inverse problems are probed on simulated and experimental data. The complexity of the simulations ranges from pure diffusion to a mixed Eley-Rideal / Langmuir-Hinshelwood mechanism for the carbon monoxide oxidation reaction. TAPsolver is bench-marked and validated against analytical solutions to show how it performs for different levels of accuracy and simulation complexities. Direct comparisons between the finite difference approximation and adjoint approach for computing derivatives are provided. A case study illustrating fitting to experimental TAP data for CO oxidation over supported platinum particles is also provided.

\subsection{TAPSolver Validation}

\subsubsection{Validation of the Forward Problem}

To confirm that TAPsolver is accurately simulating TAP reactor pulses, two examples of validation are provided: pure diffusion and irreversible adsorption. These scenarios are selected because analytical solutions are known, while exact solutions for more complex processes do not exist. 

The analytical solution for the diffusion outlet flow in a TAP reactor is 

\begin{equation}\label{diffusionAnalytical}
	 \frac{\bar{F}_{A}}{N_{pA}} = \frac{D_{eA}\pi{}}{\varepsilon{}_{b}L^{2}} \sum_{\infty{}}^{n=0}(-1)^{n}(2n+1)\exp{(-(n+0.5)^{2}\pi^{2} \frac{t D_{eA}}{\varepsilon{}_{b}L^{2}}})
\end{equation}

where $N_{pA}$ is the pulse intensity, $D_{eA}$ is the diffusion coefficient of species $A$, $\varepsilon{}_{b}$ is the void fraction of the material, $L$ is the length of the reactor and $t$ is the time \cite{gleaves1997tap}. The analytical solution for irreversible adsorption with diffusion in a TAP reactor is

\begin{equation}\label{irreversibleAnalytical}
	\frac{\bar{F}_{A}}{N_{pA}} = \frac{D_{eA}\pi{}}{\varepsilon{}_{b}L^{2}}\exp{(-k_{a}^{'})\tau{}}\sum_{\infty{}}^{n=0}(-1)^{n}(2n+1)\exp{(-(n+0.5)^{2}\pi^{2}\tau{})}
\end{equation}

where $-k_{a}^{'}$ is the rate characteristic for a unitary adsorption process and $\tau$ is the mean residence time \cite{gleaves1997tap}. Equations \ref{diffusionAnalytical} and \ref{irreversibleAnalytical} are similar and the core difference lies in the inclusion of the rate constant at the beginning of Equation \ref{irreversibleAnalytical}.

\begin{figure}[ht]
\begin{minipage}{0.5\linewidth}\label{inertAnalytical}
\includegraphics[width=\textwidth]{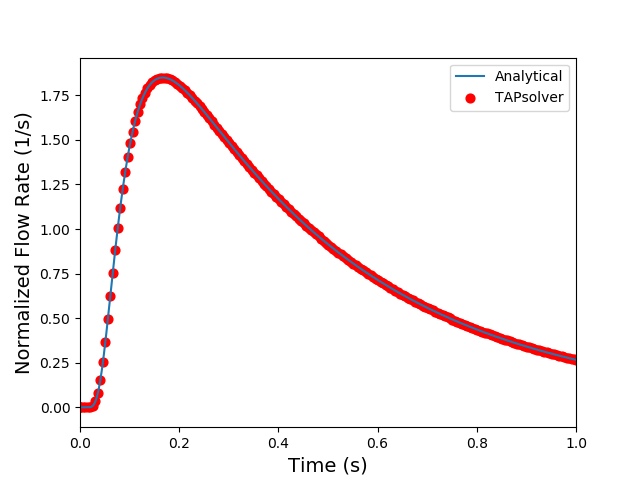}
\end{minipage}
\begin{minipage}{0.5\linewidth}
\captionsetup{justification=centering}
\includegraphics[width=\textwidth]{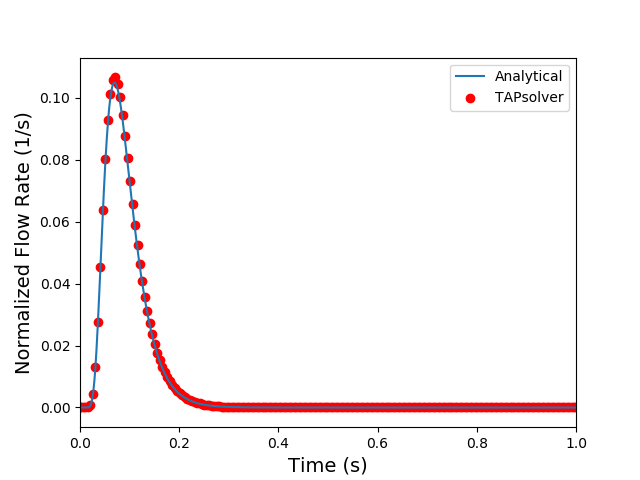}
\end{minipage}
\caption{Comparisons between the analytical solutions and TAPsolver simulations for pure Knudsen diffusion (left) and an irreversible adsorption process (right)}\label{comparingAnalytical}
\end{figure}

The left panel of Figure \ref{comparingAnalytical} shows the comparison between the analytical diffusion solution and the simulated data generated in TAPsolver, while the right panel of Figure \ref{comparingAnalytical} shows the comparison between the analytical and simulated irreversible adsorption processes. The differences between the analytical and simulated data in each of these examples are negligible, confirming the accuracy of the TAPsolver simulation. 

\subsubsection{Validation of the Inverse Problem}
To test the accuracy of the adjoint-based optimization routine implemented in TAPsolver, synthetic carbon monoxide oxidation is simulated and analyzed. The simulated mechanism included reversible carbon monoxide adsorption, irreversible dissociative $O_{2}$ adsorption, and both Eley-Rideal and Langmuir-Hinshelwood reactions that lead to the formation of carbon dioxide. These elementary reactions and their values are presented in Table \ref{lang_eley_CO}, as well as the initial guesses and the converged values following optimization.

\begin{figure}[htp]
\begin{center}
\captionsetup{justification=centering}
\subfloat A{
  \includegraphics[clip,width=0.45\columnwidth]{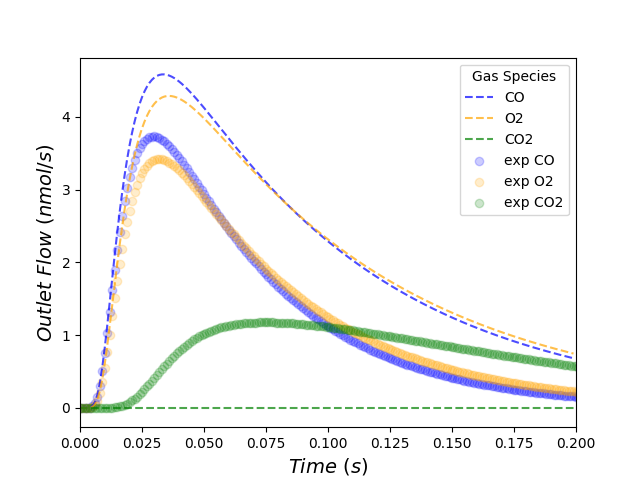}
}
\subfloat B{
  \includegraphics[clip,width=0.45\columnwidth]{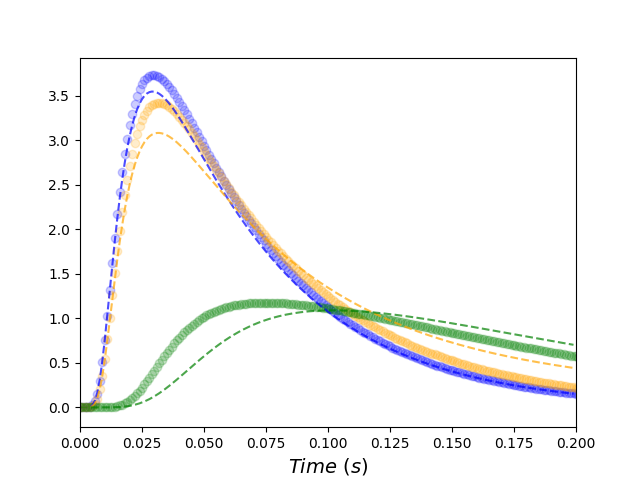}%
}
\subfloat C{
  \includegraphics[clip,width=0.45\columnwidth]{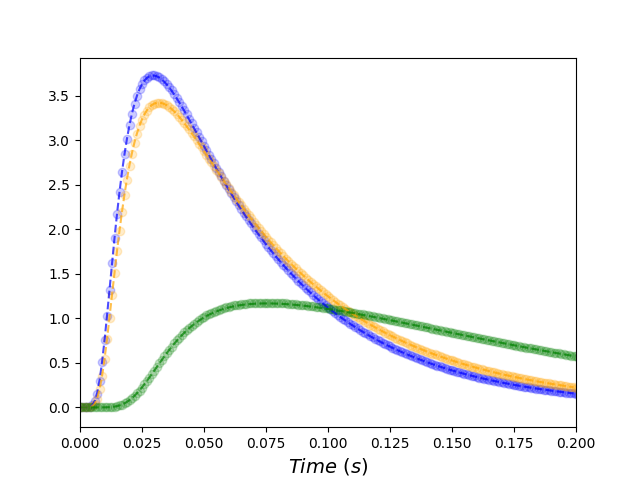}%
}
\caption{Convergence of the kinetic parameters to the synthetic values using the optimization routine in TAPsolver from the initial values (A) to the optimal values (C).}\label{convergenceSyn}
\end{center}
\end{figure}

The initial values of the kinetic parameters used for optimization are all $1.0*10^{-10}$. These values are intentionally far from the true values to illustrate the robustness of the optimization routine even when no prior knowledge of the kinetic parameters is available. Even though elementary reactions 2 and 4 were considered to be irreversible, reverse rate constants were considered during optimization. Exact matches between the fitted kinetic parameters and the synthetic values are not obtained, but strong agreement is observed with the reverse rate constants being notable exceptions. The convergence of the optimization problem is presented in Figure \ref{convergenceSyn}. With the initial guesses used, no reaction is initially observed and the reactants act as if they are purely diffusive. Gradually, the kinetic parameters are altered such that both the reactant and product curves match the synthetic data. 
This result validates that TAPSolver is able to successfully fit a curve of synthetic data, and that the primary parameters that govern the kinetics are correctly identified, even with poor initial guesses. However, not all kinetic parameters are accurately identified since the objective function is not sensitive to all parameters. This highlights an inherent limitation of fitting TAP pulses, since the data from a single pulse will not be sensitive to all parameters.

\begin{center} 
\small\addtolength{\tabcolsep}{-5pt}
\begin{longtable}{|c|c|c|c|c|c|}
\caption{List of the elementary reaction values used in the generation of the synthetic carbon monoxide oxidation data. Dashed values in the 'Actual Value' column indicate that the parameter was not included in the simulation and is therefore effectively zero.} \label{lang_eley_CO} \\

\hline Kinetic Parameter & Reaction &  Actual Value  &  Initial Guess & Converged Value & Units\\
\endfirsthead
\hline 1f & $CO + * \rightarrow CO^{*}$ & $1.5*10^{0}$  &  $1*10^{-10}$ & $1.48*10^{0}$ & $\frac{cm^{3}}{nmol s}$ \\ 
\hline 1b & $CO^{*} \rightarrow CO + *$ &  $1.5*10^{-1}$  &  $1*10^{-10}$ & $3.25*10^{-5}$ & $\frac{1}{s}$ \\
\hline 2f & $O_{2} + 2* \rightarrow 2O^{*}$ &  $5.00*10^{-3}$  & $1*10^{-10}$ & $4.70*10^{-3}$ & $\frac{cm^{6}}{nmol^{2} s}$ \\
\hline 2b & $2O^{*} \rightarrow O_{2} + 2*$ &  --  & $1*10^{-10}$ & $1.00*10^{-10}$ & $\frac{cm^{3}}{nmol s}$ \\ 
\hline 3f & $CO^{*} + O* \leftrightarrow CO_{2} + 2*$ &  $1.05*10^{1}$  &  $1*10^{-10}$ & $1.04*10^{1}$ & $\frac{cm^{3}}{nmol s}$ \\
\hline 3b & $CO_{2} + 2* \leftrightarrow CO^{*} + O*$ &  $1.5*10^{-2}$  & $1*10^{-10}$ & $6.10*10^{-3}$ & $\frac{cm^{6}}{nmol^{2} s}$  \\
\hline 4f & $CO + O* \rightarrow CO_{2} + *$ &  $2.02*10^{1}$  &  $1*10^{-10}$ & $2.50*10^{1}$ & $\frac{cm^{3}}{nmol s}$\\
\hline 4b & $CO_{2} + * \rightarrow CO + O*$ &  --  &  $1*10^{-10}$ & $5.28*10^{-3}$ & $\frac{cm^{3}}{nmol s}$ \\ 
\hline 
\end{longtable}
\end{center}

\subsection{TAPsolver Efficiency and Precision}

\begin{figure}[htp]
\begin{center}
\captionsetup{justification=centering}
\subfloat{%
  \includegraphics[clip,width=0.45\columnwidth]{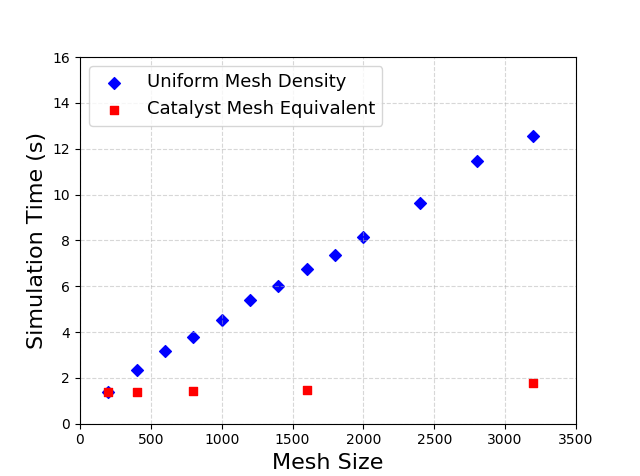}%
}
\subfloat{%
  \includegraphics[clip,width=0.45\columnwidth]{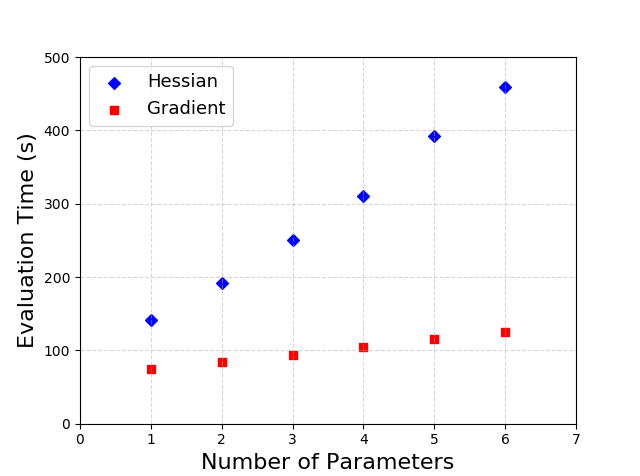}%
}
\caption{The influence of mesh density on simulation time is shown on the left, where global (Uniform Mesh Density) and local (Catalyst Mesh Equivalent) mesh refinements are compared. On the right, evaluation times for the gradient and Hessian of carbon monoxide oxidation involving one to six of the kinetic parameters shown defined in Table \ref{lang_eley_CO}. }\label{EfficiencyFigures}
\end{center}
\end{figure}

The influence of the mesh size was examined and is shown on the left in Figure \ref{EfficiencyFigures}. A uniform mesh can be utilized, where the inert and catalyst zones have identical mesh distributions. There is a linear increase in simulation time with an in increase in mesh size.

As mentioned in Section \ref{tapImplem}, an alternative method of defining the mesh has been implemented in TAPsolver, where increases in the precision of the catalyst zone can be made without refinement of the inert zones. The timing for a refined mesh is also shown in Figure  \ref{EfficiencyFigures} alongside the equivalent uniform mesh. Refinement in this context is defined as the number of times the number of cells in the catalyst zone are doubled. If the number of cells is initially four, a refinement of two will lead to a catalyst mesh size of sixteen, while a refinement of five will lead to a catalyst mesh size of one-hundred and twenty-eight.
The reduction in simulation time, even at smaller mesh sizes, by targeting the region of refinement is significant and can lead to the efficient and accurate simulation of reactors when a strong concentration profile within the catalyst is observed.
In the current implementation of TAPsolver, increasing the number of time steps leads to a linear increase in simulation time.
An explicit mixed Euler and Crank-Nicolson time stepping method is being utilized. However, future versions will explore implicit time stepping strategies  \cite{skare2012gryphon} that may be substantially more efficient, particularly in the case of stiff problems.

The time requirements for evaluating the gradient and Hessian are shown on the right of Figure \ref{EfficiencyFigures}. Once again, simulated carbon monoxide data is used as a test. A linear increase in simulation time is observed in both figures, although the slope is substantially lower for the gradient calculation. This is a considerable improvement over numerical differentiation, where the time needed to compute derivatives scales quadratically with the number of parameters.
In this example, the entire Hessian is calculated explicitly without considering symmetry, and therefore should be faster if used in optimization routines. Similar to inefficiencies in the forward problem, improvements to the efficiency of the simulator in TAPsolver will help make adjoint calculations less expensive, since the construction and evaluation of adjoints is directly dependent on the number of time steps.

The precision of the algorithmic differentiation in TAPsolver is also explored through comparisons between derivatives computed using the finite difference (FD) and algorithmic differentiation approaches. The central difference approximation \cite{gill1980computing,gill1983computing,barton1992computing} of first-order derivatives is defined as 
\begin{equation}\label{firstCentral}
    \frac{\partial f}{\partial k_{i}} = \frac{f(k_{i} + h) - f(k_{i} - h)}{2h}
\end{equation}

where $f$ is an arbitrary function of parameters $k_i$ and $h$ is a step size.

The step size is selected to be a fraction of the parameter value. To test the convergence of these values, multiple step sizes were selected, including 1/50, 1/500 and 1/5000. Comparisons between the first order derivatives calculated by the central difference and adjoint approaches at the initial guess values of each parameter are presented in Tables \ref{initialFirstDeriv} and \ref{finalFirstDeriv}.

\begin{center}
\small\addtolength{\tabcolsep}{-5pt}
\begin{longtable}{|c|c|c|c|c|}
\caption{Values of the first order derivatives found through FD approximations  and the adjoint approach at the initial guess. Multiple step sizes were used for FD to show the convergence of the solution.} \label{initialFirstDeriv} \\
\hline Kinetic \ Parameter &  1/50 & 1/500 & 1/5000 & A.D.\\
\endfirsthead
\hline 1f & $-8.71*10^{1}$ & 0.0 & 0.0
 & $-8.71*10^{1}$ \\
\hline 1b & 0.0  & 0.0 & 0.0
 & $1.04*10^{-9}$ \\
\hline 2f &  $-2.28*10^{4}$ & $-2.28*10^{4}$ & 0.0 & $-2.28*10^{4}$ \\ 
\hline 2b &  0.0  & 0.0 & 0.0 & $5.63*10^{-13}$ \\ 
\hline 3f &  0.0 & 0.0 & 0.0 & $-3.32*10^{-17}$ \\ 
\hline 3b & 0.0 & 0.0 & 0.0 & $-1.50*10^{-15}$ \\ 
\hline 4f &  0.0  & 0.0 & 0.0 & $-1.03*10^{-15}$ \\ 
\hline 4b & 0.0 & 0.0 & 0.0 & $2.00*10^{-15}$ \\ 
\hline 
\end{longtable}
\end{center}

\begin{center}
\small\addtolength{\tabcolsep}{-5pt}
\begin{longtable}{|c|c|c|c|c|}
\caption{Values of the first order derivatives found through FD approximations  and the adjoint approach at the converged value. Multiple step sizes were used for FD to show the convergence of the solution.} \label{finalFirstDeriv} \\
\hline Kinetic \ Parameter &  1/50 & 1/500 & 1/5000 & A.D.\\
\endfirsthead
\hline 1f & $2.82*10^{0}$ & $2.62*10^{0}$ & $2.60*10^{0}$
 & $2.60*10^{0}$ \\  
\hline 1b & $3.10*10^{-2}$  & $-4.43*10^{-2}$ & 0
 & $-2.88*10^{-2}$ \\ 
\hline 2f &  $-2.20*10^{3}$ & $5.16*10^{1}$ & $4.63*10^{1}$ & $4.57*10^{1}$ \\ 
\hline 2b &  $4.83*10^{-1}$  & $4.83*10^{-1}$ & $4.83*10^{-1}$ & $4.83*10^{-1}$ \\ 
\hline 3f &  $7.67*10^{-2}$ & $7.54*10^{-2}$ & $7.52*10^{-2}$ & $7.52*10^{-2}$ \\ 
\hline 3b & $-1.76*10^{0}$ & $-1.76*10^{0}$ & $-1.76*10^{0}$ & $-1.76*10^{0}$ \\ 
\hline 4f &  $-1.74*10^{-3}$  & $-1.87*10^{-3}$ & $-1.89*10^{-3}$ & $-1.89*10^{-3}$ \\ 
\hline 4b & $-1.76*10^{-1}$ & $-1.76*10^{-1}$ & $-1.76*10^{-1}$ & $-1.76*10^{-1}$ \\ 
\hline 
\end{longtable}
\end{center}
\newpage

The results in Table \ref{initialFirstDeriv} indicate that when the initial guess is far from the optimum the FD approximation routinely fails, yielding derivatives of zero for many parameters. These parameters cannot be optimized through a finite difference approach unless an improved initial guess is provided. The derivatives at the final converged solution (Table \ref{finalFirstDeriv}) indicate that the finite difference approach is significantly more accurate near the optimum, although some small discrepancies are still observed. For example, the sign of the derivative for parameter 1b changes from positive to negative depending on the step size used, highlighting a lack of stability in the finite difference approach.
The implication of this finding is that TAPsolver will be much more robust to weak initial guesses, enabling the use of less-biased initial guesses (e.g. setting all rate constants a constant value) or values from DFT that may capture trends in the rate constants but have absolute values that are off by many orders of magnitude \cite{medford2014assessing}.
In addition, the computational time required to compute algorithmic derivatives scales linearly with the number of rate constants (although the prefactor is large), suggesting that the approach will be more efficient than the FD approximation for large reaction networks. 

\subsection{Application: analyzing experimental carbon monoxide oxidation data}

\begin{figure}[bt]
\centering
\captionsetup{justification=centering}
\includegraphics[width=11cm]{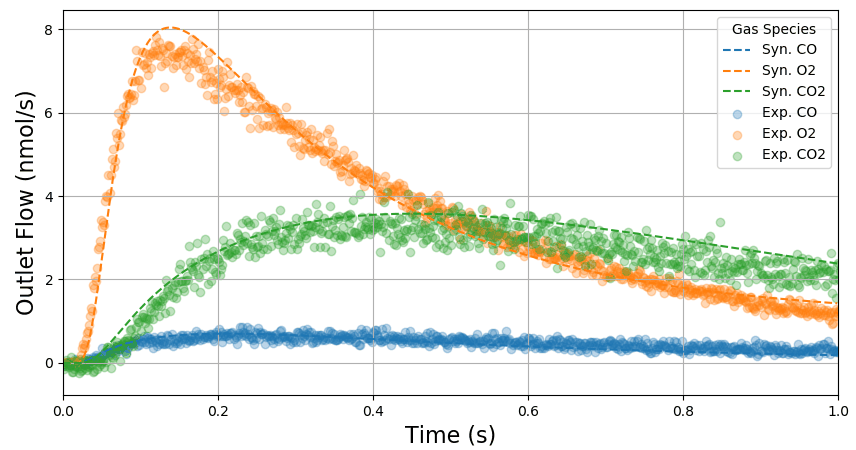}
\caption{An Eley-Rideal/Langmuir-Hinshelwood reaction mechanism fitted to experimental carbon monoxide oxidation data using TAPsolver with thermodynamic constraints.
}\label{co_oxidation_fit_eley}
\end{figure}

Carbon monoxide oxidation on platinum catalysts has frequently been studied over the last decades \cite{harold1991kinetics,harold1991kinetics,salomons2007use,herz1980surface}, including variations to reactant partial pressures \cite{farkas2013high}, cluster sizes \cite{li2013kinetic,allian2011chemisorption}, surface cleavages \cite{berlowitz1988kinetics} and support materials \cite{nibbelke1997kinetic,alayon2009highly,grass1997kinetics}. In this case study, a microkinetic model consisting of both Langmuir-Hinshelwood and Eley-Rideal mechanisms, with elementary steps shown in Table \ref{expTable} is used to fit the experimental TAP response. Thermodynamic constraints are applied to ensure that the rate constants obey the overall thermochemistry of the reaction. The resulting fit accurately captures the behavior of the TAP pulse response (Figure \ref{co_oxidation_fit_eley}), and the resulting forward and reverse rate constants for each step are provided in Table \ref{expTable}. 

This case study illustrates the power of the TAPsolver approach, yielding a full set of thermodynamically-consistent kinetic parameters from a single TAP pulse, measured in around 1 second. However, it is worth recalling that not all parameters will be well-determined, since even in the case of fitting to synthetic data it was not possible to accurately determine all kinetic parameters. Moreover, some slight discrepancies can be seen between the model and experimental data, particularly at short and long timescales for CO$_2$. This may be due to limitations in the microkinetic model used, such as neglecting adsorbate-adsorbate interactions or possible missing reactions such as oxygen transfer to the bulk. Addressing these challenges will require further development to the TAPsolver framework, such as integration with global optimization routines \cite{zhai2019data}, inclusion of additional elementary steps and lateral adsorbate interactions, or application of model reduction criteria\cite{neath2012bayesian,sakamoto1986akaike}. Nonetheless, this brief case study illustrates TAPsolver's ability to fit even noisy data sets, and highlights the ability to directly extract intrinsic kinetic parameters from TAP data.

\newpage
\begin{center}
\small\addtolength{\tabcolsep}{-5pt}
\begin{longtable}{|c|c|c|c|c|c|c|} 
\caption{List of kinetic parameters observed while fitting the Langmuir-Hinshelwood (L.H.) and Eley-Rideal (E.R.) reaction mechanisms with and without thermodynamic consistency constraints.} \label{expTable} \\
\hline  & Reaction &  L.H.  &  E.R. & L.H. Thermo. & E.R. Thermo. & Units \\ 
\endfirsthead
\hline 1f & $CO + * \rightarrow CO^{*}$ & $3.88*10^{1}$ & $7.94*10^{1}$ & $3.94*10^{1}$ & $4.3*10^{1}$ & $\frac{cm^{3}}{nmol s}$ \\
\hline 1b & $CO^{*} \rightarrow CO + *$ & $1.24*10^{1}$ & $2.89*10^{1}$ & $6.72*10^{0}$ & $1.69*10^{1}$ & $\frac{1}{s}$ \\ 
\hline 2f & $O_{2} + * \rightarrow O_{2}^{*}$ & $1.78*10^{0}$ & $1.59*10^{0}$ & $5.44*10^{-2}$ & $9.73*10^{-1}$ & $\frac{cm^{3}}{nmol s}$ \\ 
\hline 2b & $O_{2}^{*} \rightarrow O_{2} + *$ & $5.53*10^{0}$ & $1.04*10^{1}$ & $1.91*10^{-1}$ & $8.79*10^{1}$ & $\frac{1}{s}$ \\ 
\hline 3f & $O_{2}^{*} + * \rightarrow 2O{*}$ & $1.03*10^{0}$ & $1.26*10^{0}$ & $1.08*10^{1}$ & $8.68*10^{-2}$ & $\frac{cm^{3}}{nmol s}$ \\ 
\hline 3b & $2O{*} \rightarrow O_{2}^{*} + *$ & $5.21*10^{-2}$ & $2.46*10^{-2}$ & $5.23*10^{-21}$ & $8.06*10^{-4}$ & $\frac{cm^{3}}{nmol s}$ \\ 
\hline 4f & $CO^{*} + O^{*} \rightarrow CO_{2} + 2*$ & $7.16*10^{-1}$ & -- & $1.48*10^{1}$ & -- & $\frac{cm^{3}}{nmol s}$\\ 
\hline 4b & $CO_{2} + 2* \rightarrow CO^{*} + O^{*}$ & $3.19*10^{-1}$ & -- & $5.62*10^{-20}$ & -- & $\frac{cm^{3}}{nmol s}$ \\ 
\hline 5f & $CO + O* \leftrightarrow CO_{2}^{*} + *$ & -- & $6.76*10^{0}$ & -- & $5.20*10^{0}$ & $\frac{cm^{3}}{nmol s}$\\ 
\hline 5b & $CO_{2}^{*} + * \leftrightarrow CO + O*$ & -- & $3.59*10^{-1}$ & -- & $1.68*10^{0}$ & $\frac{cm^{6}}{nmol^{2} s}$ \\
\hline
\end{longtable}
\end{center}

\section{Conclusion}
The analysis of TAP reactor data, an often arduous process, has been streamlined through the development of a Python program called TAPsolver that features novel analysis methods, including algorithmic differentiation and thermodynamic consistency optimization constraints. Algorithmic differentiation provides the highest accuracy derivatives and can reduce the cost of gradient estimation, particularly when the number of parameters is large. Applying constraints to the thermodynamics can reduce the number of potential reaction mechanisms and provide physically meaningful results. These improvements make the top-down identification of microkinetic models from transient kinetic data more feasible. The forward problem has been validated through comparisons with analytical solutions and the efficiency/accuracy of the current implementation of algorithmic differentiation is demonstrated. Similarly, the inverse problem was validated by fitting noiseless, simulated carbon monoxide oxidation data. Finally, the work flow was applied to experimental carbon monoxide oxidation data on a platinum catalyst. 

The TAPsolver code provides a flexible new tool for TAP data analysis that can easily be integrated with emerging techniques in machine learning and optimization through a convenient Python interface. The tool has been developed in collaboration with experimentalists, facilitating easy transfer of parameters and data between experiment and simulation, and the code utilized input files that make it easy to explore different reaction mechanisms or change assumptions about reaction kinetics. The use of algorithmic differentiation enables accurate derivatives even far from the optimum, enabling convergence even with weak initial guesses, and gradient calculations scale linearly with the number of parameters. These advantages make TAPsolver particularly well-suited for problems with complex reaction mechanisms. We hope that TAPsolver provides a convenient platform for implementation of new TAP data analysis techniques, and lowers the barrier to TAP data analysis for both experimental and computational researchers.

\section*{Acknowledgements}
Support for this work was provided by the U.S. Department of Energy (USDOE), Office of Energy Efficiency and Renewable Energy (EERE), Advanced Manufacturing Office Next Generation R\&D Projects under contract no. DE-AC07-05ID14517. The authors are also grateful to Dr. Christian Reece for discussions of simulation methods and input file construction.

\section*{Conflict of Interest}
The authors declare no competing financial interest.

\bibliographystyle{ieeetr}
\bibliography{template}

\end{document}